\documentclass{emulateapj} 
\usepackage{apjfonts} 
\usepackage{natbib}

\shorttitle{Evolution of the $M_\star$--SFR--$Z$ Relation}
\shortauthors{Niino}

\begin{document}

\title
{The Redshift Evolution of the Relation between Stellar Mass, \\
Star Formation Rate, and Gas Metallicity of Galaxies}

\author{Yuu Niino \altaffilmark{1}}

\email{yuu.niino@nao.ac.jp}

\altaffiltext{1}{Division of Optical and IR Astronomy, 
National Astronomical Observatory of Japan, 2-21-1 Osawa, Mitaka, Tokyo, Japan}

\begin{abstract}
We investigate the relation between stellar mass ($M_\star$), star formation rate (SFR), and metallicity ($Z$) of galaxies, 
so called the fundamental metallicity relation, in the galaxy sample of the Sloan Digital Sky Survey Data Release 7. 
We separate the galaxies into narrow redshift bins and compare the relation at different redshifts, 
and find statistically significant ($> 99$\%) evolution. 
We test various observational effects that might cause seeming $Z$ evolution, 
and find it difficult to explain the evolution of the relation only by the observational effects. 
In the current sample of low redshift galaxies, 
galaxies with different $M_\star$ and SFR are sampled from different redshifts, 
and there is degeneracy between $M_\star$/SFR and redshift. 
Hence it is not straightforward to distinguish 
a relation between $Z$ and SFR from a relation between $Z$ and redshift. 
The separation of the intrinsic relation from the redshift evolution effect 
is a crucial issue to understand evolution of galaxies. 
\end{abstract}

\keywords{galaxies: abundances --- galaxies: evolution}

\section{INTRODUCTION}

Stellar mass ($M_\star$), star formation rate (SFR), and metallicity ($Z$) of galaxies
are essential parameters to understand evolution of galaxies 
\citep[e.g.,][]{Kauffmann:03a, Brinchmann:04a, Tremonti:04a, Salim:07a}. 
The relation between $M_\star$ and $Z$ has been studied at various redshifts
\citep[e.g.,][]{Tremonti:04a, Savaglio:05a, Erb:06a, Liu:08a}. 
\citet{Ellison:08a} added the 3rd parameter to the $M_\star$--$Z$ relation, 
and found the 3D correlation between $M_\star$, SFR, and $Z$  
of field star forming galaxies at low redshifts ($z\sim0.1$) in the Sloan Digital Sky Survey (SDSS) galaxy sample. 
Recently, \citet[hereafter M10]{Mannucci:10a} and \citet{Lara-Lopez:10a} showed 
that high redshift galaxies ($z\lesssim3$) agree with the extrapolation 
of the $M_\star$--SFR--$Z$ relation defined at low redshifts (so called the fundamental metallicity relation). 

The no-evolution of the $M_\star$--SFR--$Z$ relation with redshift 
suggests existence of a physical process 
which affects evolution of galaxies in the wide range of redshift behind the relation. 
Some models are already proposed to explain the origin of the $M_\star$--SFR--$Z$ relation 
\citep{Dave:12a,Dayal:12a,Yates:12a}

Although the correlation between $M_\star$, SFR, and $Z$ is tested at various redshifts 
\citep[M10;][]{Lara-Lopez:10a, Richard:11a, Yabe:11a, Nakajima:12a, Cresci:12a, Wuyts:12a}, 
galaxies observed at different redshifts typically have different $M_\star$ and SFR. 
Hence the comparison of galaxy metallicities at different redshifts 
in a same range of $M_\star$ and SFR is hardly done. 
Thus it is currently difficult to distinguish a fundamental relation 
between $M_\star$, SFR, and $Z$ that stands in wide range of redshift, 
from a series of the $M_\star$--$Z$ relations at various redshifts. 

In this study, we separate the low redshift galaxy sample ($z<0.3$) 
in which the $M_\star$--SFR--$Z$ relation is defined into narrow redshift bins, 
and investigate redshift evolution of the relation. 
Throughout this paper, we assume the fiducial cosmology 
with $\Omega_{\Lambda}=0.7$, $\Omega_{m}=0.3$, and $H_0=$ 70 km s$^{-1}$ Mpc$^{-1}$.

\section{THE GALAXY SAMPLE}
\label{sec:sample}

We draw galaxies for our analysis from 
the SDSS MPA-JHU Data Release 7 catalog\footnote{http://www.mpa-garching.mpg.de/SDSS/DR7/} 
(hereafter the MPA-JHU catalog), 
with similar selection criteria to that in M10 as follows. 
We select galaxies (1) at redshifts between 0.07 and 0.3, 
(2) the signal-to-noise ratio (S/N) of H$\alpha$ line $>25$, 
(3) $A_V < 2.5$, (4) H$\alpha$ to H$\beta$ flux ratio (H$\alpha$/H$\beta$) $>2.5$, 
and (5) not strongly affected by active galactic nuclei (AGN) according to \citet{Kauffmann:03b}.
These selection criteria are identical to those in M10. 

The target galaxy selection method for the spectroscopic observation in the SDSS provides 
highly uniform and complete sample of galaxies above applied magnitude limit 
that is $r$-band Petrosian magnitude $m_{r,{\rm Pet}}\leq17.77$ \citep{Strauss:02a}. 
However, galaxies fainter than this limit which have been selected by different selection methods 
\citep[e.g. luminous red galaxy sample,][]{Eisenstein:01a} are also included in the MPA-JHU catalog. 
To avoid possible selection effect in those galaxies, 
we impose following selection criteria in addition to the M10 criteria. 
Galaxies in our sample (6) have $m_{r,{\rm Pet}}\leq17.77$, 
(7) are selected for spectroscopic targets as galaxy candidates \citep{Stoughton:02a}, 
and (8) spectroscopically confirmed as galaxies by the SDSS spectroscopic pipeline 
(see the SDSS DR7 website\footnote{http://www.sdss.org/dr7/} for detail). 
Finally $\sim 110000$ galaxies are left in our sample. 
We note that $\sim$ 90\% of the galaxies in our sample is at $z<0.15$, 
although we collect galaxies at redshifts up to 0.3. 

We use $M_\star$ and SFR of galaxies listed in the MPA-JHU catalog 
\citep{Kauffmann:03a, Brinchmann:04a, Salim:07a}, 
and measure metallicity of the galaxies using empirical method by \citet{Maiolino:08a}. 
Following M10, we use $R_{23}=$ ([OII]$\lambda$3727+[OIII]$\lambda$4958,5007)/H$\beta$, 
and [NII]$\lambda$6584/H$\alpha$ as the indicators of metallicity. 
The line flux ratios are corrected for extinctions 
using H$\alpha$/H$\beta$ and the extinction curve given by \citet{Cardelli:89a}. 
When both indicators are applicable to a galaxy, 
we use mean of the 12 + log(O/H) values given by the two indicators, 
as far as the two values agree with each other within 0.25 dex. 
When the two indicators disagree, we remove the galaxy from our sample. 
Only 3\% of galaxies in our sample is rejected here, similarly to the case in M10. 

\section{CORRELATION OF SFR AND METALLICITY}
\label{sec:result}

\begin{figure}
\begin{center}
\includegraphics[scale=0.5]{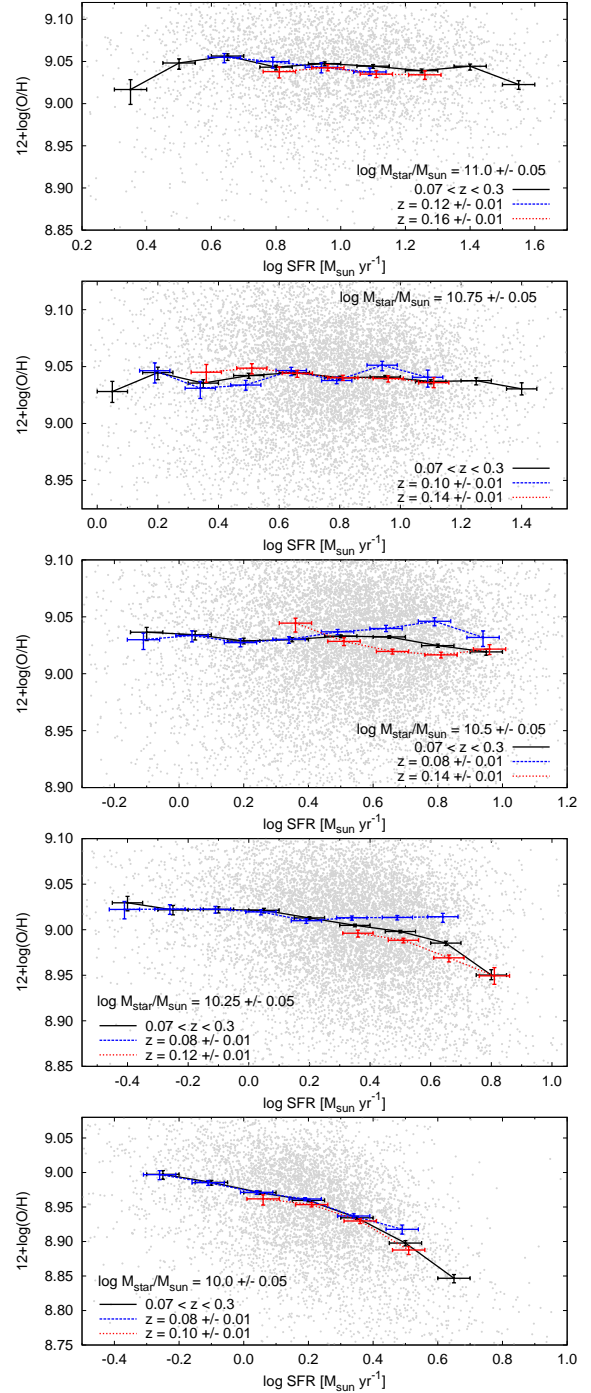}
\caption {
The correlation between SFR and $Z$ of the galaxies 
with log~$M_\star/M_\odot = 11.0\pm0.05$, $10.75\pm0.05$, $10.5\pm0.05$, $10.25\pm0.05$, and $10.0\pm0.05$ 
from top to bottom, respectively. 
The data points connected with solid lines (black) represents 
the median metallicity of the galaxies in the whole range of redshift $0.07<z<0.3$ in each SFR bin. 
The data points connected with dashed (blue) and dotted (red) lines 
represent the median metallicities of the low and high redshift samples. 
The vertical error bars represent the error of mean 
of the metallicity distribution in each bin
(the scatter of the sample divided by the square root of the sample number). 
The horizontal error bars represent the width of the SFR bin. 
The low and high redshift data points are slightly shifted sideway for visibility. 
The grey dots represent individual galaxies. 
A colored version of the figure is available in the online journal. 
}
\label{fig:SFRZ}
\end{center}
\end{figure}

We show the correlation between SFR and metallicity of the galaxies in mass ranges 
log~$M_\star/M_\odot = 10.0\pm0.05$, $10.25\pm0.05$, $10.5\pm0.05$, $10.75\pm0.05$, and $11.0\pm0.05$ in figure~\ref{fig:SFRZ}. 
We separate the galaxies into SFR bins with $\Delta$log~SFR = 0.1 dex, 
and measure median of metallicities in each bin. 
The correlation between SFR and metallicity is not clear 
among the galaxies with log~$M_\star/M_\odot \gtrsim 10.5$ (the top three panels). 
On the other hand, among the galaxies with log~$M_\star/M_\odot \lesssim 10.25$, 
higher SFR galaxies have lower metallicity than lower SFR galaxies (the bottom two panels), 
consistently to the results of M10. 

\begin{figure}
\begin{center}
\includegraphics[scale=0.5]{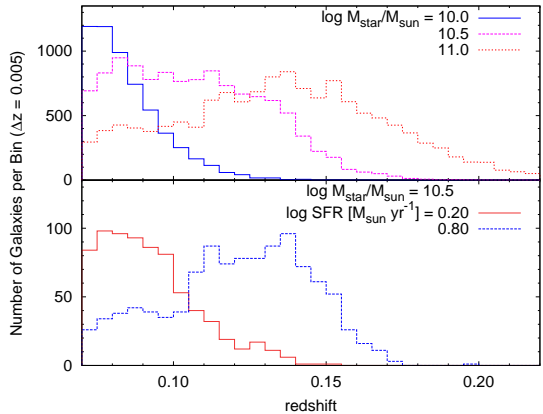}
\caption {
{\it Top panel}: the redshift distribution of the galaxies in our sample 
with stellar mass: log~$M_\star/M_\odot = 10.0\pm0.05$, $10.5\pm0.05$, and $11.0\pm0.05$. 
{\it Bottom panel}: the redshift distributions of the log~$M_\star/M_\odot = 10.5$ galaxies, 
with log~SFR [$M_\odot$yr$^{-1}$] = $0.2\pm0.05$ and $0.8\pm0.05$. 
}
\label{fig:zdist}
\end{center}
\end{figure}

We also separate the galaxies into redshift bins with $\Delta z$ = 0.02, 
and compare the SFR--$Z$ relations at different redshifts (figure~\ref{fig:SFRZ}). 
The plotted redshift bins are different in different mass ranges, 
due to the difference of redshift distribution 
of the galaxies in each mass range (the top panel of figure~\ref{fig:zdist}). 
Each data point in figure~\ref{fig:SFRZ} contains more than 50 galaxies. 
Although systematic difference of the SFR--$Z$ relation at different redshifts 
is not clear in the mass ranges log~$M_\star/M_\odot \gtrsim 10.75$, 
the galaxies with with log~$M_\star/M_\odot \lesssim 10.5$ 
show redshift evolution of the SFR--$Z$ relation. 

In the mass ranges log~$M_\star/M_\odot \lesssim 10.5$, 
galaxies with higher SFR shows larger chemical evolution. 
Making the negative correlation between SFR and $Z$ steeper at higher redshift. 
The low redshift sample of galaxies with log~$M_\star/M_\odot = 10.5$ 
shows a positive correlation between SFR and $Z$ 
which is consistent to the results of \citet{Yates:12a}, 
while the high redshift galaxies with the same mass range 
shows the negative correlation similarly to the case in lower mass ranges. 

As discussed in \citet{Brisbin:12a},  
the extent of the metallicity evolution with redshift shown in figure~\ref{fig:SFRZ} 
is small ($\lesssim 0.05$ dex) due to the narrow redshift range we can investigate with our sample. 
However, most datapoints which show the redshift evolution in figure~\ref{fig:SFRZ} 
have the error of mean $< 0.01$ dex indicating that the evolution is statistically significant. 
We discuss statistical significance of the evolution of the relation further in \S~\ref{sec:KStest}. 

It should be noted that a combination effect of fiber covering fraction 
and metallicity gradient in a galaxy may cause seeming metallicity evolution with redshift. 
In the SDSS, galaxy spectra are obtained only in the 3 arcsec fiber aperture. 
The fiber aperture can contain larger area of a target galaxy at higher redshifts, 
while galaxies tend to have lower metallicity in their outskirt than at their center \citep[e.g.,][]{Zaritsky:94a}. 
We discuss this issue in \S~\ref{sec:fiber}. 
There are also other observational effects that may affect 
the redshift evolution of the $M_\star$--SFR--$Z$ relation. 
The limiting magnitude of the SDSS spectroscopic target selection $m_{r,{\rm Pet}}\leq17.77$
and/or the H$\alpha$ S/N $> 25$ threshold in our sample selection 
may cause some redshift dependent sampling bias. 
The larger noise in galaxy spectra at higher redshift is 
also a possible source of artificial redshift effect. 
We discuss these effects in \S~\ref{sec:Llimit} and \S~\ref{sec:composite}, respectively. 

In figure~\ref{fig:SFRZ}, the galaxies with log~$M_\star/M_\odot \gtrsim 10.5$ 
and log~SFR [$M_\odot$yr$^{-1}$] $\lesssim 0.5$ 
show possible negative evolution of metallicity (higher metallicity at higher redshift), 
although the error of mean is large (the sample size is small). 
One possible explanation for the negative chemical evolution of the low SFR galaxies 
is decrease of SFR in galaxies with high SFR and low metallicity. 
When SFR and metallicity are negatively correlated, 
some of high-SFR low-metallicity galaxies may decrease their SFR 
before significant chemical evolution decreasing the median metallicity of low SFR galaxies, 
while others continue star formation undergoing chemical enrichment 
and increasing the median metallicity of high SFR galaxies. 

\begin{figure}
\begin{center}
\includegraphics[scale=0.5]{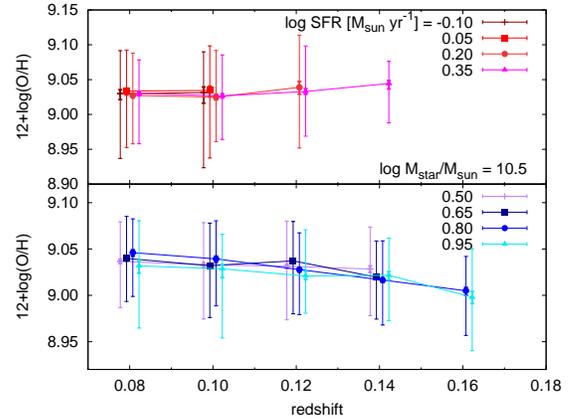}
\caption {
The metallicity evolution 
of the galaxies with log~$M_\star/M_\odot = 10.5\pm0.05$. 
The data points represent the median, the 1$\sigma$ scatter (thin error bars), 
and the error of mean (thick error bars) of the galaxy metallicities. 
The width of each bin is 0.02 dex and 0.1 dex for redshift and SFR, respectively. 
The data points are slightly shifted sideway for visibility. 
}
\label{fig:redshift_evo}
\end{center}
\end{figure}

To investigate the redshift evolution further, 
we plot the median metallicities of the galaxies with log~$M_\star/M_\odot = 10.5\pm0.05$ 
and various SFR as functions of redshift in figure~\ref{fig:redshift_evo}. 
The metallicities of the high SFR galaxies (log~SFR [$M_\odot$yr$^{-1}$] $\gtrsim 0.5$) is lower at higher redshifts, 
while the metallicities of the low SFR galaxies shows negative or no evolution, 
confirming the results in figure~\ref{fig:SFRZ}. 

In the bottom panel of figure~\ref{fig:zdist}, 
we show redshift distributions of the galaxies with log~$M_\star/M_\odot = 10.5$, 
and log~SFR [$M_\odot$yr$^{-1}$] $= 0.2$ \& 0.8. 
The galaxies with different SFR are sampled from different redshifts, 
as well as in the case of different $M_\star$ shown in the top panel. 
In figure~\ref{fig:SFRZ}, the SFR--$Z$ relations for the whole redshift range ($0.07<z<0.3$) 
follows the low redshift relation in the low SFR end, 
while metallicity of the high SFR galaxies follows the high redshift relation, 
consistently to the difference of the redshift distributions between the low and high SFR galaxies. 

\section{DISCUSSION}
\subsection{Kolmogorov-Smirnov Test}
\label{sec:KStest}

\begin{figure}
\begin{center}
\includegraphics[scale=0.5]{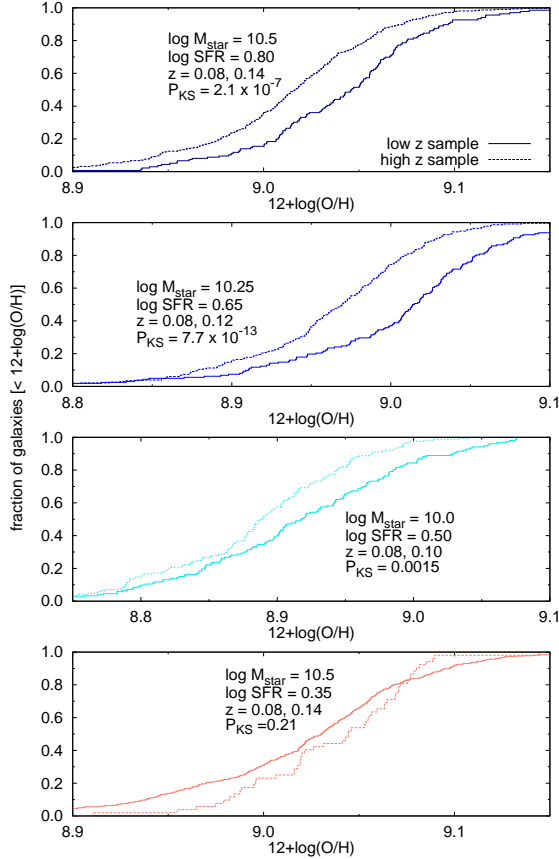}
\caption {
The cumulative metallicity distributions of the galaxies 
with the four sets of $M_\star$ and SFR, at low and high redshifts. 
The top 3 panels show the high SFR galaxies whose metallicity is lower at higher redshift, 
while the bottom panel shows the low SFR galaxies whose metallicity is (possibly) higher at higher redshift. 
In each panel, solid and dashed histograms represent low and high redshift samples, respectively. 
The plotted redshifts are different for different set of $M_\star$ and SFR, 
due to the difference of the redshift distributions of the galaxies 
with different $M_\star$ and SFR (see figure~\ref{fig:zdist}).
}
\label{fig:KStest}
\end{center}
\end{figure}

To examine the significance of the redshift evolution we find in \S~\ref{sec:result}, 
we perform the two-sample Kolmogorov-Smirnov (KS) test, 
which calculates the probability that two samples 
can be drawn from a same probability distribution function. 
In figure~\ref{fig:KStest}, we show metallicity distributions 
of the galaxies with the four sets of $M_\star$ and SFR, 
which show notable redshift evolution in figure~\ref{fig:SFRZ}, 
at two different redshifts each. 

The positive metallicity evolution (lower-metallicity at higher-redshift) 
in the mass ranges log~$M_\star/M_\odot \lesssim 10.5$ is statistically significant 
to high confidence level $>99$\% (the top 3 panels of figure~\ref{fig:KStest}). 
The statistical significance of the evolution of the galaxies 
with log~$M_\star/M_\odot \sim 10.25$ and 10.5 
is very high with the KS test probability $P_{\rm KS} = 7.7\times10^{-13}$ and $2.1\times10^{-7}$. 

Although the statistical significance of the evolution of the galaxies with log~$M_\star/M_\odot = 10.0$ 
is lower than in the cases of the galaxies with log~$M_\star/M_\odot \sim 10.25$ and 10.5 
due to the very narrow range of redshift where we can investigate log~$M_\star/M_\odot = 10.0$ galaxies, 
the evolution in this mass range is also larger than statistical error ($P_{\rm KS} = 1.5\times10^{-3}$). 
The extent of the evolution of the galaxies with log~$M_\star/M_\odot = 10.0$ 
is similar to that of the log~$M_\star/M_\odot = 10.5$ galaxies ($\sim 0.03$ dex), 
although the redshift range investigated is 3 times smaller. 
This suggests that less massive galaxies are more rapidly evolving in metallicity, 
consistently to the findings of previous studies 
\citep[e.g.,][but see also \citeauthor{Lamareille:09a} \citeyear{Lamareille:09a}]{Savaglio:05a}. 

The possible negative metallicity evolution of the high-mass, low-SFR galaxies is statistically less significant 
than the positive evolution of the high SFR galaxies (the bottom panel of figure~\ref{fig:KStest}), 
as expected from the large error of mean in figure~\ref{fig:SFRZ}. 

\subsection{The Effect of Fiber Covering Fraction}
\label{sec:fiber}

The combination effect of fiber covering fraction and metallicity gradient in a galaxy 
may cause seeming metallicity evolution with redshift. 
The SDSS spectroscopic fiber (3 arcsec) contains larger area of a target galaxy at higher redshift, 
and galaxies tend to have lower metallicity in their outskirt than at their center \citep[e.g.,][]{Zaritsky:94a}. 

\begin{figure}
\begin{center}
\includegraphics[scale=0.5]{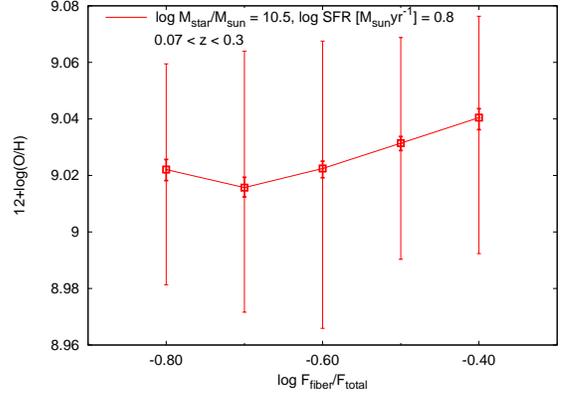}
\caption {
The median, the 1$\sigma$ scatter, and the error of mean of the galaxy metallicities 
with log~$M_\star/M_\odot = 10.5$ and log~SFR [$M_\odot$yr$^{-1}$] $= 0.8$, 
as a function of the fiber covering fraction. 
Here the fiber covering fraction is the ratio of flux in fiber 
to total flux of the object measured in $r$-band. 
The width of the fiber covering fraction bins is 0.1 dex. 
}
\label{fig:fiber}
\end{center}
\end{figure}

We show the relation between the fiber covering fraction (the fiber to total flux ratio in $r$-band)
and the derived metallicity of the galaxies with log~$M_\star/M_\odot = 10.5$ 
and log~SFR [$M_\odot$yr$^{-1}$] $= 0.8$ in figure~\ref{fig:fiber}. 
The galaxies with larger covering fraction have higher-metallicity, 
contrary to what is expected from the metallicity gradient. 
This trend may be a result of the negative correlation 
between galaxy radius and metallicity found by \citet{Ellison:08a}. 
The absence of the expected negative correlation 
between the fiber covering fraction and the metallicity suggests 
that the fiber covering fraction is determined by intrinsic size of galaxies rather than their redshift. 
We note that the galaxies have significant variation of radus (6--15 kpc, Petrosian) even in the narrow bin of $M_\star$ and SFR. 

It should be noted that the gradient effect may be hidden in figure~\ref{fig:fiber} 
canceled by the effect of the radius--metallicity correlation, 
and hence the absence of the gradient effect in the plot doesn't completely rule out the effect. 
However, if the redshift evolution of metallicity 
results from the effect of the metallicity gradient, 
we expect larger evolution for galaxies with larger radius and the gradient. 
Considering that galaxies with larger $M_\star$ tend to have larger radius, 
and dwarf galaxies have flatter metallicity gradient than spirals \citep[e.g.,][]{Lee:06a}, 
the finding of the metallicity evolution only in high specific SFR galaxies 
is opposite to the expected trend in the case of the gradient effect. 
Hence we consider it is difficult to explain the metallicity evolution 
found in this study only by the effect of the metallicity gradient. 

\subsection{Continuum and H$\alpha$ Line Luminosity Limits}
\label{sec:Llimit}

As mentioned in \S~\ref{sec:sample}, 
the SDSS spectroscopic target galaxies are magnitude limited $m_{r,{\rm Pet}}\leq17.77$, 
and we select galaxies with H$\alpha$ S/N $> 25$ for this study. 
Hence the broad band limiting magnitude and/or the H$\alpha$ flux limit 
may cause some redshift dependent sampling bias. 

\begin{figure}
\begin{center}
\includegraphics[scale=0.5]{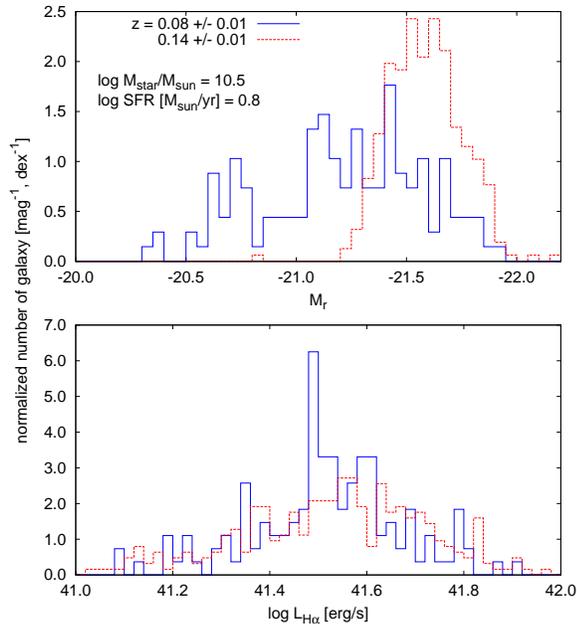}
\caption {
{\it Top panel}: the $r$-band luminosity distribution of the galaxies 
with log~$M_\star/M_\odot = 10.5$ and log~SFR [$M_\odot$yr$^{-1}$] $= 0.8$ 
sampled from the two different redshifts. 
{\it Bottom panel}: same as the top panel, but the H$\alpha$ luminosity distribution. 
}
\label{fig:lumdist}
\end{center}
\end{figure}

In figure~\ref{fig:lumdist}, we plot $M_r$ and H$\alpha$ luminosity distributions of the galaxies 
with log~$M_\star/M_\odot = 10.5$ and log~SFR [$M_\odot$yr$^{-1}$] $= 0.8$ at redshift 0.08 and 0.14.
With the fixed $M_\star$ and SFR, the galaxies at $z = 0.14$ 
are clearly brighter than the $z = 0.08$ galaxies in $r$-band, 
suggesting that the $r$-band limiting magnitude plays an important role in this sample. 
On the other hand, the H$\alpha$ luminosity distributions at the two different redshifts are not largely different. 
We note that galaxies with log~SFR [$M_\odot$yr$^{-1}$] $= 0.8$ 
typically have H$\alpha$ S/N $\sim 60$ at $z=0.14$, 
and hence few galaxies in this SFR range are rejected 
by the H$\alpha$ S/N limit by the H$\alpha$ S/N threshold. 

\begin{figure*}
\begin{center}
\includegraphics[scale=0.5]{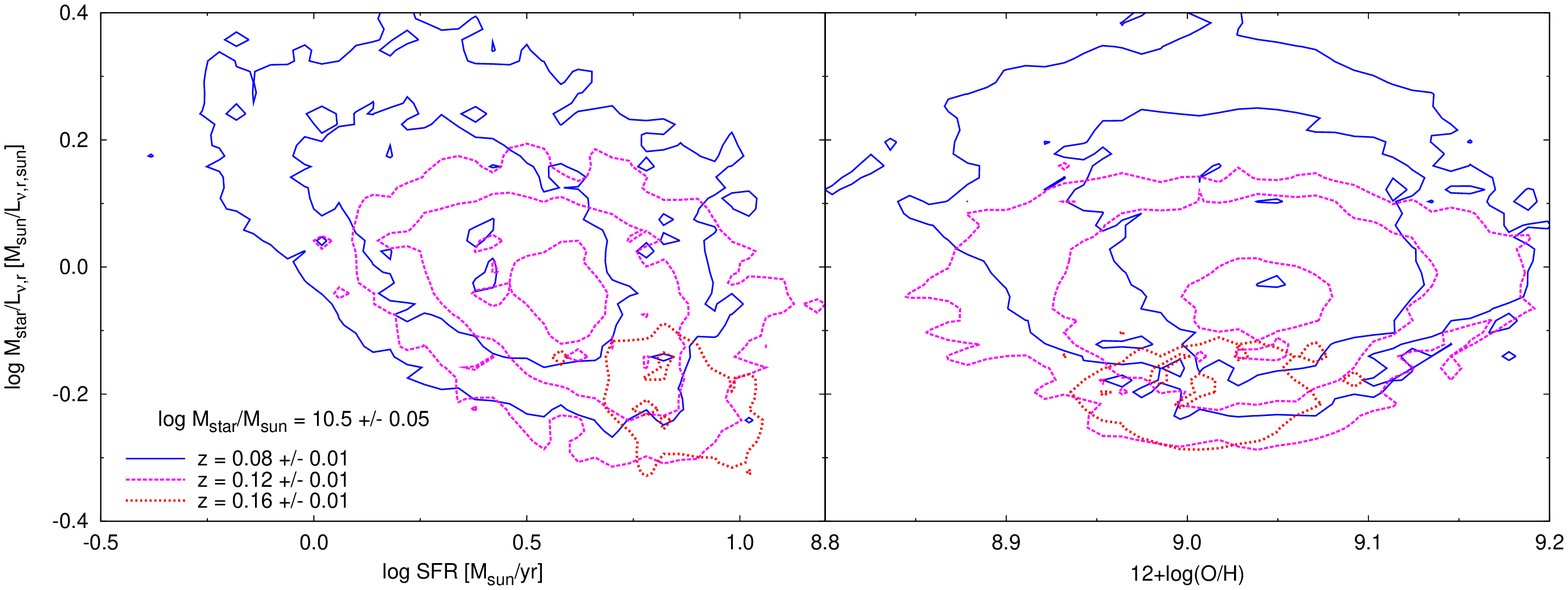}
\caption {
{\it Left panel}: the distribution of the galaxies with log~$M_\star/M_\odot = 10.5$, 
on the SFR versus mass-to-luminosity ratio plane. 
The distributions of the galaxies at $z = 0.08, 0.12$ and $0.16 \pm 0.01$ 
are shown with solid (blue), dashed (magenta), and dotted (red) contours, respectively. 
The contours indicates $d^2n_{\rm gal}/d{\rm log SFR}d{\rm log}(M_\star/L_{\nu,r}) = 1, 3, 9\ldots$ [dex$^{-2}$]. 
The $r$-band solar luminosity is $L_{\nu,r,\odot} = 5.9\times10^{18}$ erg sec$^{-1}$ Hz$^{-1}$ \citep{Blanton:03a}. 
{\it Right panel}: same as the left panel, but on the metallicity versus mass-to-luminosity ratio plane. 
}
\label{fig:masslum}
\end{center}
\end{figure*}

To discuss the effect of the limiting magnitude, 
we plot distributions of the galaxies with log~$M_\star/M_\odot = 10.5$ 
in some redshift bins on the SFR versus $r$-band mass-to-luminosity ratio ($M_\star/L_{\nu,r}$) plane 
and the metallicity versus $M_\star/L_{\nu,r}$ plane in figure~\ref{fig:masslum}. 

In the left panel of figure~\ref{fig:masslum}, $M_\star/L_{\nu,r}$ of the galaxies clearly correlates with SFR, 
causing the difference of redshift distribution between the galaxy samples 
with different SFR (the bottom panel of figure~\ref{fig:zdist}). 
On the other hand, no clear correlation between metallicity and $M_\star/L_{\nu,r}$ is seen in the right panel. 
Furthermore, it is also notable that galaxies at higher redshifts 
have systematically lower metallicity than lower redshift galaxies with similar $M_\star/L_{\nu,r}$. 

\begin{figure}
\begin{center}
\includegraphics[scale=0.5]{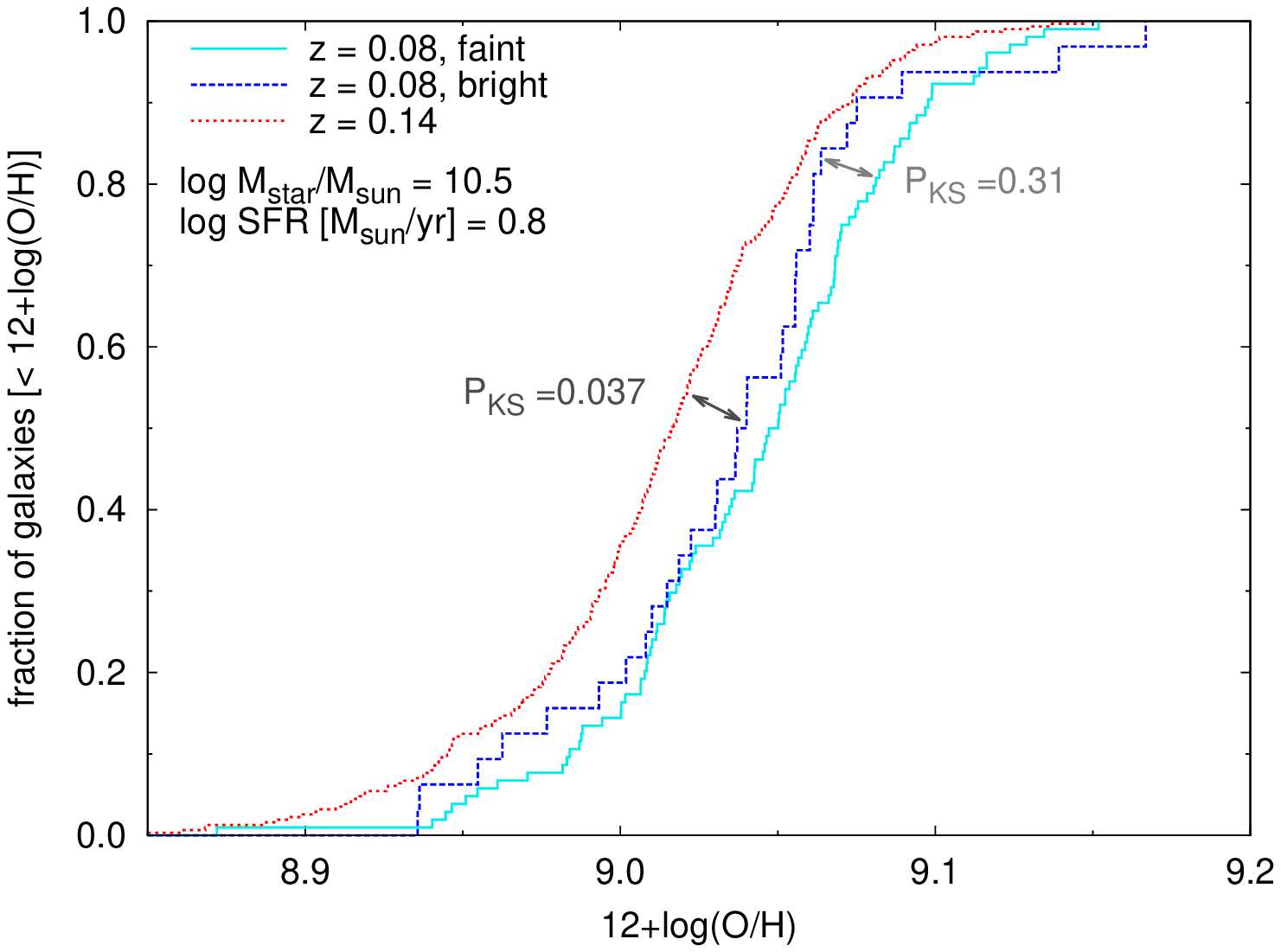}
\caption {
The cumulative metallicity distributions of the galaxies with log~$M_\star/M_\odot = 10.5$ 
and log~SFR [$M_\odot$yr$^{-1}$] $= 0.8$, at the two different redshifts. 
The low redshift sample is separated into the bright ($M_r < -21.5$) and the faint ($M_r \geq 21.5$) subsamples. 
The galaxies in the bright sample have similar  $M_r$ to that of the high redshift sample, 
where faint galaxies are not selected as spectroscopic targets. 
The KS test probabilities between the distributions are indicated together. 
}
\label{fig:maglimit_KS}
\end{center}
\end{figure}

To test the effect of the limiting magnitude further, 
we separate the galaxies with log~$M_\star/M_\odot = 10.5$ 
and log~SFR [$M_\odot$yr$^{-1}$] $= 0.8$ at redshift 0.08 into two subsamples. 
One is bright ($M_r < -21.5$), and the other is faint ($M_r \geq 21.5$). 
We compare the metallicity distributions of the bright/faint samples at $z = 0.08$ 
and the sample at $z = 0.14$ in figure~\ref{fig:maglimit_KS}. 
The galaxies in the bright sample at $z = 0.08$ have similar $M_r$ to the galaxies at $z = 0.14$. 
Although the bright sample is small, the KS test between the bright sample and the $z = 0.14$ sample 
indicates the metallicity distributions of the two sample is significantly different ($P_{\rm KS} = 0.037$), 
while the faint sample and the bright sample is broadly consistent ($P_{\rm KS} = 0.31$). 
Hence it is unlikely that the limiting magnitude effect 
is the primary source of the redshift evolution of the $M_\star$--SFR--$Z$ relation. 

\subsection{Noise Effect and Mean Spectra}
\label{sec:composite}

With similar $M_\star$ and SFR, spectra of high redshift galaxies 
would have lower S/N than those of low redshift galaxies. 
This expected trend possibly affect the metallicity estimate in a redshift dependent way. 
We also note that figure~\ref{fig:lumdist} suggests 
that high redshift galaxies have smaller H$\alpha$ equivalent width than low redshift galaxies, 
and hence it is possible that high redshift galaxies suffer more 
from uncertainties of the continuum subtraction than low redshift galaxies. 
To reduce the noise effect, we compose mean spectra of the galaxies 
with log~$M_\star/M_\odot = 10.5\pm0.05$ and log~SFR [$M_\odot$yr$^{-1}$] $= 0.8\pm0.05$ 
at redshifts $0.08\pm0.01,\ 0.10\pm0.01,\ 0.12\pm0.01,\ 0.14\pm0.01$, and $0.16\pm0.01$, 
and measure [NII]$\lambda$6584/H$\alpha$ and $R_{23}$ of the mean spectrum at each redshift. 

We average spectra collected from the SDSS data archive server\footnote{http://das.sdss.org/www/html/} 
after shifting wavelength to the center of each redshift bin. 
To reduce the effect of stellar absorption lines on the emission line measurement, 
we fit continuum of the mean spectra with stellar spectral energy distribution (SED) models, 
and subtract the stellar component models form the mean spectra. 
We perform the SED fitting with the {\it SEDfit} software package \citep{Sawicki:12a, Sawicki:98a} 
which utilize the population synthesis models of \citet{Bruzual:03a}. 
We examine five cases of star formation history: 
simple stellar population, constant star formation, 
and exponentially decaying star formation with $\tau =$ 0.2, 1.0, and 5.0 Gyr 
assuming stellar metallicity to be $Z_\odot$. 

\begin{figure*}
\begin{center}
\includegraphics[scale=0.5]{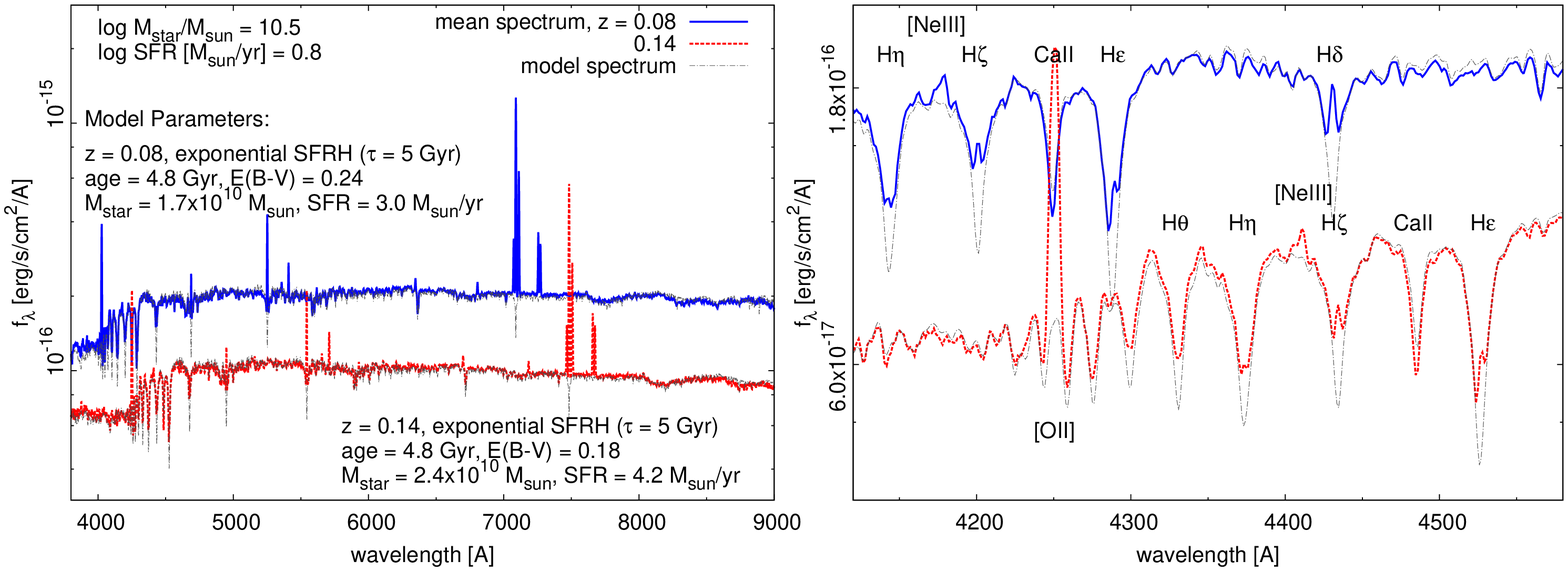}
\caption {
{\it Left panel}: the mean spectra of the galaxies with log~$M_\star/M_\odot = 10.5$ 
and log~SFR [$M_\odot$yr$^{-1}$] $= 0.8$, at $z=$ 0.08 and 0.14. 
The model stellar spectra are plotted together with the model parameters indicated. 
{\it Right panel}: a close up view of high order Balmer lines in the mean spectra shown in the left panel. 
}
\label{fig:composite}
\end{center}
\end{figure*}

Two mean spectra and their best fit stellar SED models 
are shown in the left panel of figure~\ref{fig:composite}. 
The mean spectra achieves very high S/N (typically $\sim$ 100 in the highest redshift bin). 
The best fit parameters of the stellar SED models indicate smaller $M_\star$ and SFR 
than log~$M_\star/M_\odot = 10.5$ and log~SFR [$M_\odot$yr$^{-1}$] $= 0.8$, 
but we note that the spectra only represents stellar populations in the spectroscopic fiber. 
In the right panel of figure~\ref{fig:composite}, 
we show a close up view of high order Balmer lines in the mean spectra. 
Although the center of each absorption line is affected by the corresponding emission line, 
the stellar SED models reproduce Balmer absorption lines in their wings. 

\begin{figure*}
\begin{center}
\includegraphics[scale=0.8]{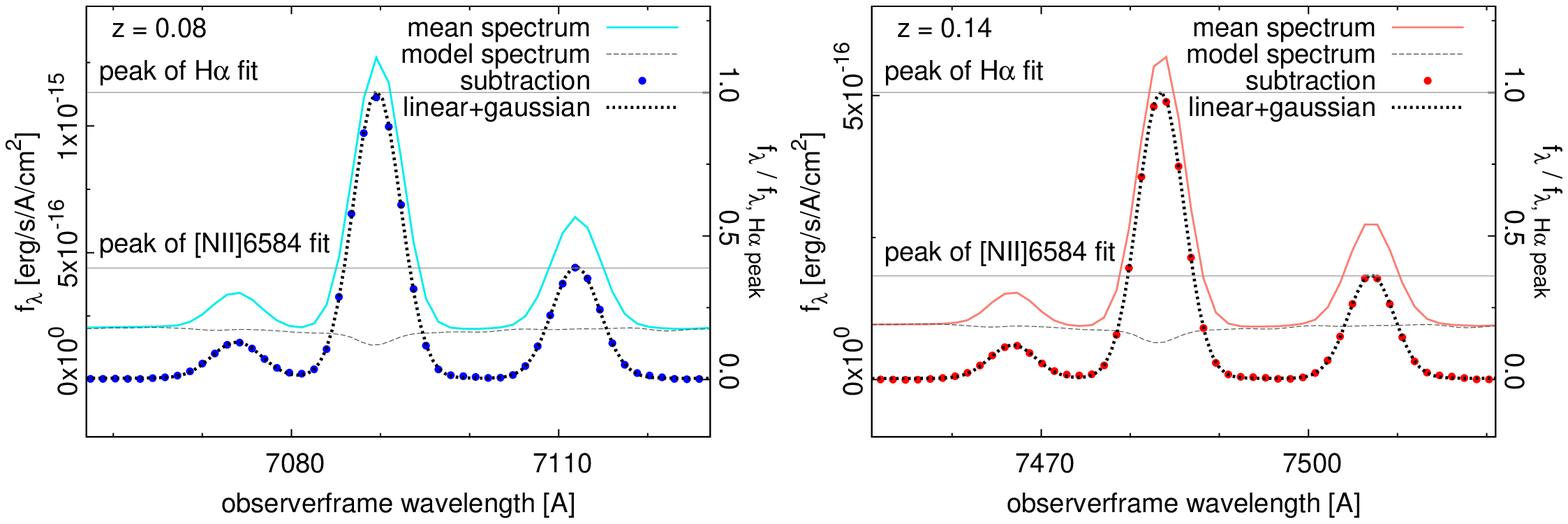}
\caption {
[NII] and H$\alpha$ lines in the mean spectra at $z=$ 0.08 and 0.14. 
Solid and dashed curves are the mean spectra 
and the model spectra which are shown in figure~\ref{fig:composite}. 
The model subtracted spectra are plotted as data points (filled circle), 
to which we perform the linear continuum plus gaussian fitting. 
Note that the noise level is too small to plot. 
The results of the line fittings are shown with dotted lines. 
The peaks of the gaussian models of H$\alpha$ 
and [NII]$\lambda$6584 lines are indicated with horizontal lines. 
The right vertical axis in each panel indicates $f_\lambda$ normalized 
to the peak of the gaussian fit of H$\alpha$ line. 
}
\label{fig:NIIHa_fit}
\end{center}
\end{figure*}

\begin{figure*}
\begin{center}
\includegraphics[scale=0.8]{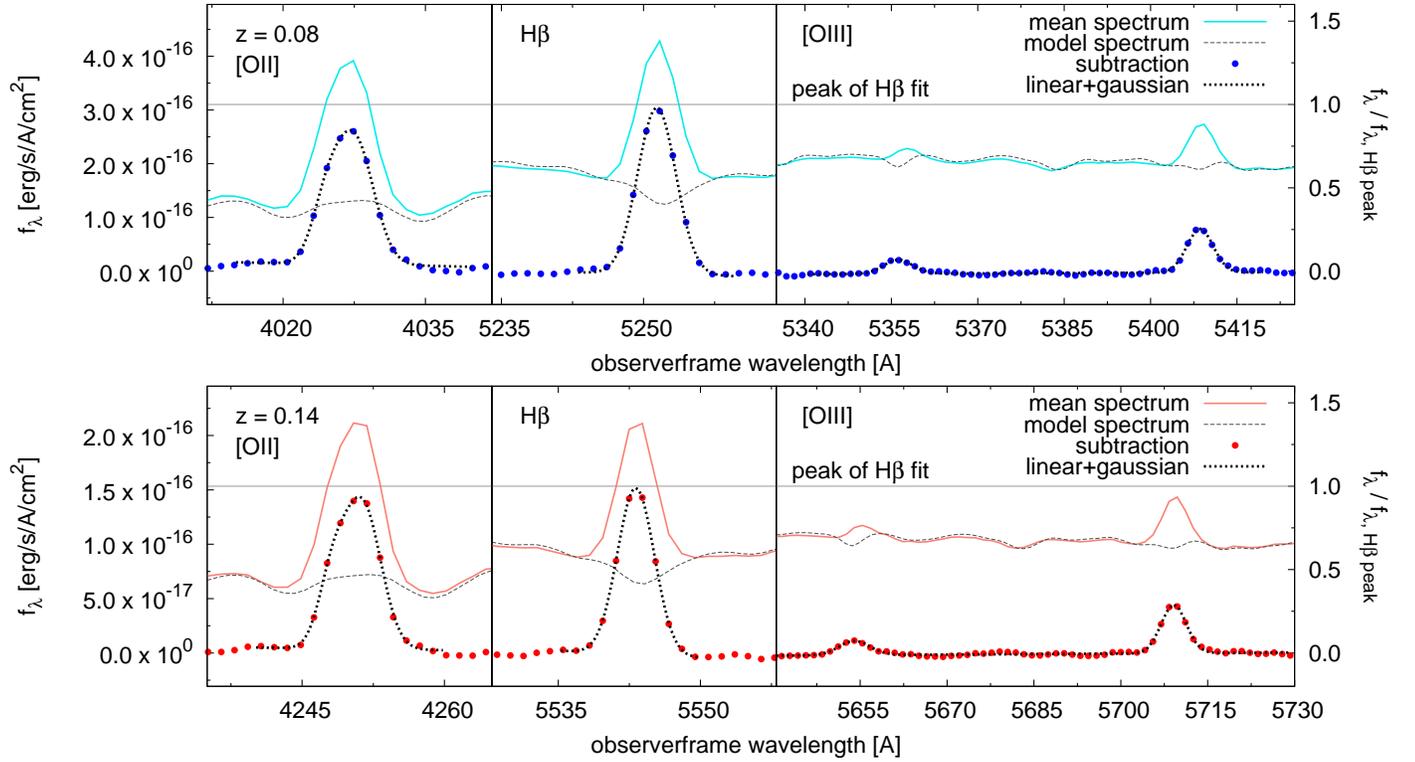}
\caption {
Same as figure~\ref{fig:NIIHa_fit}, but for $R_{23}$ lines. 
The right vertical axis indicates $f_\lambda$ normalized to the peak of H$\beta$ line. 
}
\label{fig:R23_fit}
\end{center}
\end{figure*}

After subtracting the stellar SED models, 
we fit each line with linear continuum plus single gaussian to measure flux of each line 
(we fit [OII]$\lambda$3727 doublet with linear continuum plus double gaussian). 
We show the line fittings at $z=$ 0.08 and 0.14 in figure~\ref{fig:NIIHa_fit} and \ref{fig:R23_fit}. 
When normalized to the peaks of the gaussian models of H$\alpha$, 
the spectrum at $z=0.08$ has slightly stronger [NII]$\lambda$6584 
than the spectrum at $z=0.14$ (figure~\ref{fig:NIIHa_fit}). 
We also fit [NII]$\lambda$6548 to avoid 
that the line affects the estimate of the residual continuum, 
although we don't use the line as a metallicity indicator. 
One can also find that [OII]$\lambda$3727 is slightly weaker at $z=0.08$ 
than at $z=0.14$ (figure~\ref{fig:R23_fit}), contrary to the case of [NII]$\lambda$6584. 
We note that the continuum level underneath [OII]$\lambda$3727 may be affected 
by high order Balmer emission lines whose peaks are not resolved in the mean spectra, 
and it may affect $R_{23}$, while [NII]/H$\alpha$ is free from this effect. 

\begin{figure}
\begin{center}
\includegraphics[scale=0.5]{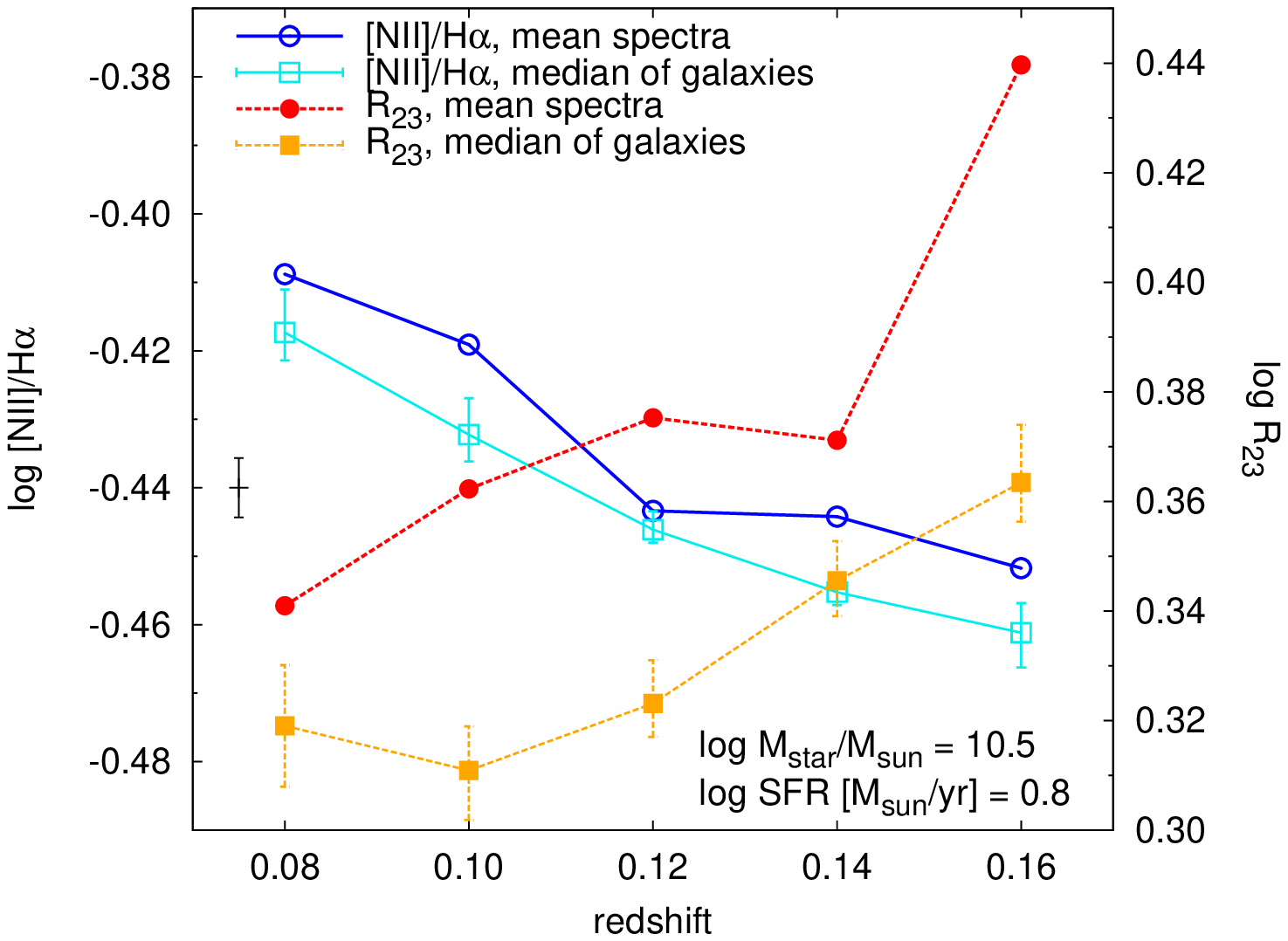}
\caption {
The redshift evolution of the metal indicating line ratios of the mean spectra. 
The median line ratios of the galaxies 
with the same $M_\star$, SFR, and redshift are plotted together. 
Each line is detected with S/N = 100--1000 in the mean spectra, 
where the vertical error bar in the left corresponds to S/N = 100. 
The vertical error bars of the median line ratios indicate the error of mean. 
}
\label{fig:LRevolution}
\end{center}
\end{figure}

The line fluxes obtained by the gaussian fit are listed in table~\ref{tb:lines} 
after corrected for extinction using H$\alpha$/H$\beta$ ratio and the \citet{Cardelli:89a} extinction curve. 
The redshift evolution of the metal indicating line ratios is plotted in figure~\ref{fig:LRevolution}, 
together with the median line ratios of the galaxies 
with the same $M_\star$, SFR, and redshift to those of the mean spectra. 
Note that the line fluxes of the mean spectra are obtained by our gaussian fit, 
while the median line ratios are obtained using the line fluxes listed in the MPA-JHU catalog. 

Galaxies at higher redshifts show smaller [NII]/H$\alpha$ and larger $R_{23}$, in their mean spectra. 
The relations between the line ratios and $Z$ are $d{\rm log([NII]/H\alpha)}/d{\rm log}Z = 0.81$ 
and $d{\rm log}R_{23}/d{\rm log}Z = -1.5$ at 12 + log(O/H) = 9.0 \citep{Maiolino:08a}, 
and hence the evolution of the line ratios in figure~\ref{fig:LRevolution} 
is broadly consistent to the metallicity evolution found in \S~\ref{sec:result}. 
The median line ratios also evolve with similar manner to the line ratios of the mean spectra, 
although they indicate smaller metal-to-hydrogen line ratio than the results with the mean spectrum, 
possibly due to the difference of spectrum fitting method. 
Thus we conclude that the metallicity evolution is not a result of noise effect. 

\begin{deluxetable*}{ccccccc}
\tablecolumns{7}
\tablewidth{0pc}
\tablecaption{The Emission Lines in the Mean Spectra}
\tablehead{
  \colhead{redshift} & 
  \colhead{[OII]$\lambda$3727} & \colhead{H$\beta$} & 
  \colhead{[OIII]$\lambda$4959} & \colhead{[OIII]$\lambda$5007} & 
  \colhead{H$\alpha$} & \colhead{[NII]$\lambda$6584} 
}
\startdata
      0.08 & $2.72\times10^{-14}$ & $1.50\times10^{-14}$ & $1.50\times10^{-15}$ & $4.15\times10^{-15}$ & $3.86\times10^{-14}$ & $1.51\times10^{-14}$ \cr
      0.10 & $1.91\times10^{-14}$ & $9.98\times10^{-15}$ & $9.95\times10^{-16}$ & $2.88\times10^{-15}$ & $2.57\times10^{-14}$ & $9.80\times10^{-15}$ \cr
      0.12 & $1.34\times10^{-14}$ & $6.86\times10^{-15}$ & $7.32\times10^{-16}$ & $2.10\times10^{-15}$ & $1.78\times10^{-14}$ & $6.40\times10^{-15}$ \cr
      0.14 & $9.18\times10^{-15}$ & $4.76\times10^{-15}$ & $5.38\times10^{-16}$ & $1.46\times10^{-15}$ & $1.24\times10^{-14}$ & $4.46\times10^{-15}$ \cr
      0.16 & $8.14\times10^{-15}$ & $3.58\times10^{-15}$ & $4.37\times10^{-16}$ & $1.27\times10^{-15}$ & $9.36\times10^{-15}$ & $3.31\times10^{-15}$ 
\enddata
\tablecomments{
  The extinction corrected line fluxes of the mean spectra of the galaxies with log~$M_\star/M_\odot = 10.5\pm0.05$ 
  and log~SFR [$M_\odot$yr$^{-1}$] $= 0.8\pm0.05$. The units are [erg s$^{-1}$ cm$^{-2}$]. 
  }  \label{tb:lines}
\end{deluxetable*}

\section{SUMMARY}

We have investigated the evolution of the $M_\star$--SFR--$Z$ relation at $z<0.3$. 
Although the redshift range we investigated is narrow, 
we found metallicity of the high SFR galaxies (log~SFR [$M_\odot$yr$^{-1}$] $\gtrsim 0.5$) 
with log~$M_\star/M_\odot \lesssim 10.5$ is evolving with $>99$\% statistical significance. 
We have examined the observational effects which may cause seeming evolution of metallicity: 
the fiber aperture effect, the sampling bias by the limiting magnitude, and the noise in the spectra. 
We found it is difficult to explain the evolution of the $M_\star$--SFR--$Z$ relation 
only by the observational effects, although some effects are not completely ruled out. 

In the current galaxy sample at low redshifts, 
galaxies with different $M_\star$ and SFR are sampled with different redshift distributions, 
and there is a degeneracy between $M_\star$/SFR, and redshift. 
Hence it is difficult to clearly separate SFR dependence of metallicity from redshift dependence. 
The metallicity evolution of galaxies with same $M_\star$ and SFR at $z<0.3$ 
found in this study suggests that the currently known $M_\star$--SFR--$Z$ relation 
arises from a combination of the intrinsic relation and the redshift evolution effect. 
Previous studies \citep[e.g., M10;][]{Lara-Lopez:10a} showed 
that $z>0.5$ galaxies agree with the extrapolation of the relation defined by $z<0.3$ galaxies, 
suggesting that the relation is not evolving with redshift. 
However, if the redshift evolution effect is included in the relation at $z<0.3$, 
the agreements of high redshift galaxies don't necessarily mean that the relation is not evolving. 
The agreements may result from the extrapolation of the redshift evolution effect. 

We need to investigate metallicity distribution of galaxies 
with same $M_\star$ and SFR at different redshifts, 
to distinguish the intrinsic relation from the redshift evolution effect. 
However, it is currently possible only for narrow range of $M_\star$, SFR, and redshift. 
The extent of the evolution we find in the current sample is very small 
due to the very narrow range of redshift we can investigate. 
Hence it is difficult to robustly exclude possibility of systematic effects in our results. 

Deeper wide field spectroscopic survey which covers wider range of redshift 
is necessary to test the evolution effect more robustly. 
For example, to study similar range of $M_\star$ and SFR at $z\sim1$
with a comparable sample number (or a survey volume, $\gtrsim$ 10\%) to this study, 
the limiting magnitude must be 5 magnitude deeper than the SDSS sample ($m \lesssim 23$),  
and the survey field must be $\gtrsim$ 20 deg$^2$ (0.3\% of the SDSS DR7 field). 
Note that the comoving volume element is $\sim$ 37 times larger at $z=1$ than at $z=0.1$. 
Future spectroscopic surveys, 
such as Subaru Measurement of Images and Redshifts (SuMIRe) project, 
will achieve such requirements. 
At redshifts $z<1$, the limiting magnitude can be brighter but the survey field must be wider. 
The robust separation of the intrinsic $M_\star$--SFR--$Z$ relation from its evolution 
will open a new way to understand galaxy evolution. 

\acknowledgments
We are grateful to the MPA/JHU group for making their galaxy catalog public.  
We thank K. Yabe for helpful discussions. 
Thanks are also due to our referee, M. A. Strauss, 
whose thoughtful comments largely improved this paper. 


\begin{thebibliography}{31}
\expandafter\ifx\csname natexlab\endcsname\relax\def\natexlab#1{#1}\fi

\bibitem[{{Blanton} {et~al.}(2003){Blanton}, {Hogg}, {Bahcall}, {Brinkmann},
  {Britton}, {Connolly}, {Csabai}, {Fukugita}, {Loveday}, {Meiksin}, {Munn},
  {Nichol}, {Okamura}, {Quinn}, {Schneider}, {Shimasaku}, {Strauss}, {Tegmark},
  {Vogeley}, \& {Weinberg}}]{Blanton:03a}
{Blanton}, M.~R., {et~al.} 2003, \apj, 592, 819

\bibitem[{{Brinchmann} {et~al.}(2004){Brinchmann}, {Charlot}, {White},
  {Tremonti}, {Kauffmann}, {Heckman}, \& {Brinkmann}}]{Brinchmann:04a}
{Brinchmann}, J., {Charlot}, S., {White}, S.~D.~M., {Tremonti}, C.,
  {Kauffmann}, G., {Heckman}, T., \& {Brinkmann}, J. 2004, \mnras, 351, 1151

\bibitem[{{Brisbin} \& {Harwit}(2012)}]{Brisbin:12a}
{Brisbin}, D., \& {Harwit}, M. 2012, \apj, 750, 142

\bibitem[{{Bruzual} \& {Charlot}(2003)}]{Bruzual:03a}
{Bruzual}, G., \& {Charlot}, S. 2003, \mnras, 344, 1000

\bibitem[{{Cardelli} {et~al.}(1989){Cardelli}, {Clayton}, \&
  {Mathis}}]{Cardelli:89a}
{Cardelli}, J.~A., {Clayton}, G.~C., \& {Mathis}, J.~S. 1989, \apj, 345, 245

\bibitem[{{Cresci} {et~al.}(2012){Cresci}, {Mannucci}, {Sommariva}, {Maiolino},
  {Marconi}, \& {Brusa}}]{Cresci:12a}
{Cresci}, G., {Mannucci}, F., {Sommariva}, V., {Maiolino}, R., {Marconi}, A.,
  \& {Brusa}, M. 2012, \mnras, 421, 262

\bibitem[{{Dav{\'e}} {et~al.}(2012){Dav{\'e}}, {Finlator}, \&
  {Oppenheimer}}]{Dave:12a}
{Dav{\'e}}, R., {Finlator}, K., \& {Oppenheimer}, B.~D. 2012, \mnras, 421, 98

\bibitem[{{Dayal} {et~al.}(2012){Dayal}, {Ferrara}, \& {Dunlop}}]{Dayal:12a}
{Dayal}, P., {Ferrara}, A., \& {Dunlop}, J.~S. 2012, arXiv:1202.4770

\bibitem[{{Eisenstein} {et~al.}(2001){Eisenstein}, {Annis}, {Gunn}, {Szalay},
  {Connolly}, {Nichol}, {Bahcall}, {Bernardi}, {Burles}, {Castander},
  {Fukugita}, {Hogg}, {Ivezi{\'c}}, {Knapp}, {Lupton}, {Narayanan}, {Postman},
  {Reichart}, {Richmond}, {Schneider}, {Schlegel}, {Strauss}, {SubbaRao},
  {Tucker}, {Vanden Berk}, {Vogeley}, {Weinberg}, \& {Yanny}}]{Eisenstein:01a}
{Eisenstein}, D.~J., {et~al.} 2001, \aj, 122, 2267

\bibitem[{{Ellison} {et~al.}(2008){Ellison}, {Patton}, {Simard}, \&
  {McConnachie}}]{Ellison:08a}
{Ellison}, S.~L., {Patton}, D.~R., {Simard}, L., \& {McConnachie}, A.~W. 2008,
  \apjl, 672, L107

\bibitem[{{Erb} {et~al.}(2006){Erb}, {Shapley}, {Pettini}, {Steidel}, {Reddy},
  \& {Adelberger}}]{Erb:06a}
{Erb}, D.~K., {Shapley}, A.~E., {Pettini}, M., {Steidel}, C.~C., {Reddy},
  N.~A., \& {Adelberger}, K.~L. 2006, \apj, 644, 813

\bibitem[{{Kauffmann} {et~al.}(2003{\natexlab{a}}){Kauffmann}, {Heckman},
  {White}, {Charlot}, {Tremonti}, {Brinchmann}, {Bruzual}, {Peng}, {Seibert},
  {Bernardi}, {Blanton}, {Brinkmann}, {Castander}, {Cs{\'a}bai}, {Fukugita},
  {Ivezic}, {Munn}, {Nichol}, {Padmanabhan}, {Thakar}, {Weinberg}, \&
  {York}}]{Kauffmann:03a}
{Kauffmann}, G., {et~al.} 2003{\natexlab{a}}, \mnras, 341, 33

\bibitem[{{Kauffmann} {et~al.}(2003{\natexlab{b}}){Kauffmann}, {Heckman},
  {Tremonti}, {Brinchmann}, {Charlot}, {White}, {Ridgway}, {Brinkmann},
  {Fukugita}, {Hall}, {Ivezi{\'c}}, {Richards}, \& {Schneider}}]{Kauffmann:03b}
---. 2003{\natexlab{b}}, \mnras, 346, 1055

\bibitem[{{Lamareille} {et~al.}(2009){Lamareille}, {Brinchmann}, {Contini},
  {Walcher}, {Charlot}, {P{\'e}rez-Montero}, {Zamorani}, {Pozzetti},
  {Bolzonella}, {Garilli}, {Paltani}, {Bongiorno}, {Le F{\`e}vre}, {Bottini},
  {Le Brun}, {Maccagni}, {Scaramella}, {Scodeggio}, {Tresse}, {Vettolani},
  {Zanichelli}, {Adami}, {Arnouts}, {Bardelli}, {Cappi}, {Ciliegi}, {Foucaud},
  {Franzetti}, {Gavignaud}, {Guzzo}, {Ilbert}, {Iovino}, {McCracken}, {Marano},
  {Marinoni}, {Mazure}, {Meneux}, {Merighi}, {Pell{\`o}}, {Pollo}, {Radovich},
  {Vergani}, {Zucca}, {Romano}, {Grado}, \& {Limatola}}]{Lamareille:09a}
{Lamareille}, F., {et~al.} 2009, \aap, 495, 53

\bibitem[{{Lara-L{\'o}pez} {et~al.}(2010){Lara-L{\'o}pez}, {Cepa},
  {Bongiovanni}, {P{\'e}rez Garc{\'{\i}}a}, {Ederoclite}, {Casta{\~n}eda},
  {Fern{\'a}ndez Lorenzo}, {Povi{\'c}}, \&
  {S{\'a}nchez-Portal}}]{Lara-Lopez:10a}
{Lara-L{\'o}pez}, M.~A., {et~al.} 2010, \aap, 521, L53

\bibitem[{{Lee} {et~al.}(2006){Lee}, {Skillman}, \& {Venn}}]{Lee:06a}
{Lee}, H., {Skillman}, E.~D., \& {Venn}, K.~A. 2006, \apj, 642, 813

\bibitem[{{Liu} {et~al.}(2008){Liu}, {Shapley}, {Coil}, {Brinchmann}, \&
  {Ma}}]{Liu:08a}
{Liu}, X., {Shapley}, A.~E., {Coil}, A.~L., {Brinchmann}, J., \& {Ma}, C.-P.
  2008, \apj, 678, 758

\bibitem[{{Maiolino} {et~al.}(2008){Maiolino}, {Nagao}, {Grazian}, {Cocchia},
  {Marconi}, {Mannucci}, {Cimatti}, {Pipino}, {Ballero}, {Calura}, {Chiappini},
  {Fontana}, {Granato}, {Matteucci}, {Pastorini}, {Pentericci}, {Risaliti},
  {Salvati}, \& {Silva}}]{Maiolino:08a}
{Maiolino}, R., {et~al.} 2008, \aap, 488, 463

\bibitem[{{Mannucci} {et~al.}(2010){Mannucci}, {Cresci}, {Maiolino}, {Marconi},
  \& {Gnerucci}}]{Mannucci:10a}
{Mannucci}, F., {Cresci}, G., {Maiolino}, R., {Marconi}, A., \& {Gnerucci}, A.
  2010, \mnras, 408, 2115

\bibitem[{{Nakajima} {et~al.}(2012){Nakajima}, {Ouchi}, {Shimasaku}, {Ono},
  {Lee}, {Foucaud}, {Ly}, {Dale}, {Salim}, {Finn}, {Almaini}, \&
  {Okamura}}]{Nakajima:12a}
{Nakajima}, K., {et~al.} 2012, \apj, 745, 12

\bibitem[{{Richard} {et~al.}(2011){Richard}, {Jones}, {Ellis}, {Stark},
  {Livermore}, \& {Swinbank}}]{Richard:11a}
{Richard}, J., {Jones}, T., {Ellis}, R., {Stark}, D.~P., {Livermore}, R., \&
  {Swinbank}, M. 2011, \mnras, 413, 643

\bibitem[{{Salim} {et~al.}(2007){Salim}, {Rich}, {Charlot}, {Brinchmann},
  {Johnson}, {Schiminovich}, {Seibert}, {Mallery}, {Heckman}, {Forster},
  {Friedman}, {Martin}, {Morrissey}, {Neff}, {Small}, {Wyder}, {Bianchi},
  {Donas}, {Lee}, {Madore}, {Milliard}, {Szalay}, {Welsh}, \& {Yi}}]{Salim:07a}
{Salim}, S., {et~al.} 2007, \apjs, 173, 267

\bibitem[{{Savaglio} {et~al.}(2005){Savaglio}, {Glazebrook}, {Le Borgne},
  {Juneau}, {Abraham}, {Chen}, {Crampton}, {McCarthy}, {Carlberg}, {Marzke},
  {Roth}, {J{\o}rgensen}, \& {Murowinski}}]{Savaglio:05a}
{Savaglio}, S., {et~al.} 2005, \apj, 635, 260

\bibitem[Sawicki(2012)]{Sawicki:12a} 
 {Sawicki}, M. 2012, accepted for publication in \pasp\ (arXiv:1210.0285)

\bibitem[{{Sawicki} \& {Yee}(1998)}]{Sawicki:98a}
 {Sawicki}, M., \& {Yee}, H.~K.~C. 1998, \aj, 115, 1329

\bibitem[{{Stoughton} {et~al.}(2002){Stoughton}, {Lupton}, {Bernardi},
  {Blanton}, {Burles}, {Castander}, {Connolly}, {Eisenstein}, {Frieman},
  {Hennessy}, {Hindsley}, {Ivezi{\'c}}, {Kent}, {Kunszt}, {Lee}, {Meiksin},
  {Munn}, {Newberg}, {Nichol}, {Nicinski}, {Pier}, {Richards}, {Richmond},
  {Schlegel}, {Smith}, {Strauss}, {SubbaRao}, {Szalay}, {Thakar}, {Tucker},
  {Vanden Berk}, {Yanny}, {Adelman}, {Anderson}, {Anderson}, {Annis},
  {Bahcall}, {Bakken}, {Bartelmann}, {Bastian}, {Bauer}, {Berman},
  {B{\"o}hringer}, {Boroski}, {Bracker}, {Briegel}, {Briggs}, {Brinkmann},
  {Brunner}, {Carey}, {Carr}, {Chen}, {Christian}, {Colestock}, {Crocker},
  {Csabai}, {Czarapata}, {Dalcanton}, {Davidsen}, {Davis}, {Dehnen},
  {Dodelson}, {Doi}, {Dombeck}, {Donahue}, {Ellman}, {Elms}, {Evans}, {Eyer},
  {Fan}, {Federwitz}, {Friedman}, {Fukugita}, {Gal}, {Gillespie}, {Glazebrook},
  {Gray}, {Grebel}, {Greenawalt}, {Greene}, {Gunn}, {de Haas}, {Haiman},
  {Haldeman}, {Hall}, {Hamabe}, {Hansen}, {Harris}, {Harris}, {Harvanek},
  {Hawley}, {Hayes}, {Heckman}, {Helmi}, {Henden}, {Hogan}, {Hogg}, {Holmgren},
  {Holtzman}, {Huang}, {Hull}, {Ichikawa}, {Ichikawa}, {Johnston}, {Kauffmann},
  {Kim}, {Kimball}, {Kinney}, {Klaene}, {Kleinman}, {Klypin}, {Knapp},
  {Korienek}, {Krolik}, {Kron}, {Krzesi{\'n}ski}, {Lamb}, {Leger},
  {Limmongkol}, {Lindenmeyer}, {Long}, {Loomis}, {Loveday}, {MacKinnon},
  {Mannery}, {Mantsch}, {Margon}, {McGehee}, {McKay}, {McLean}, {Menou},
  {Merelli}, {Mo}, {Monet}, {Nakamura}, {Narayanan}, {Nash}, {Neilsen},
  {Newman}, {Nitta}, {Odenkirchen}, {Okada}, {Okamura}, {Ostriker}, {Owen},
  {Pauls}, {Peoples}, {Peterson}, {Petravick}, {Pope}, {Pordes}, {Postman},
  {Prosapio}, {Quinn}, {Rechenmacher}, {Rivetta}, {Rix}, {Rockosi}, {Rosner},
  {Ruthmansdorfer}, {Sandford}, {Schneider}, {Scranton}, {Sekiguchi}, {Sergey},
  {Sheth}, {Shimasaku}, {Smee}, {Snedden}, {Stebbins}, {Stubbs}, {Szapudi},
  {Szkody}, {Szokoly}, {Tabachnik}, {Tsvetanov}, {Uomoto}, {Vogeley}, {Voges},
  {Waddell}, {Walterbos}, {Wang}, {Watanabe}, {Weinberg}, {White}, {White},
  {Wilhite}, {Wolfe}, {Yasuda}, {York}, {Zehavi}, \& {Zheng}}]{Stoughton:02a}
{Stoughton}, C., {et~al.} 2002, \aj, 123, 485

\bibitem[{{Strauss} {et~al.}(2002){Strauss}, {Weinberg}, {Lupton}, {Narayanan},
  {Annis}, {Bernardi}, {Blanton}, {Burles}, {Connolly}, {Dalcanton}, {Doi},
  {Eisenstein}, {Frieman}, {Fukugita}, {Gunn}, {Ivezi{\'c}}, {Kent}, {Kim},
  {Knapp}, {Kron}, {Munn}, {Newberg}, {Nichol}, {Okamura}, {Quinn}, {Richmond},
  {Schlegel}, {Shimasaku}, {SubbaRao}, {Szalay}, {Vanden Berk}, {Vogeley},
  {Yanny}, {Yasuda}, {York}, \& {Zehavi}}]{Strauss:02a}
{Strauss}, M.~A., {et~al.} 2002, \aj, 124, 1810

\bibitem[{{Tremonti} {et~al.}(2004){Tremonti}, {Heckman}, {Kauffmann},
  {Brinchmann}, {Charlot}, {White}, {Seibert}, {Peng}, {Schlegel}, {Uomoto},
  {Fukugita}, \& {Brinkmann}}]{Tremonti:04a}
{Tremonti}, C.~A., {et~al.} 2004, \apj, 613, 898

\bibitem[{{Wuyts} {et~al.}(2012){Wuyts}, {Rigby}, {Sharon}, \&
  {Gladders}}]{Wuyts:12a}
{Wuyts}, E., {Rigby}, J.~R., {Sharon}, K., \& {Gladders}, M.~D. 2012,
  arXiv:1202.5267

\bibitem[{{Yabe} {et~al.}(2011){Yabe}, {Ohta}, {Iwamuro}, {Yuma}, {Akiyama},
  {Tamura}, {Kimura}, {Takato}, {Moritani}, {Sumiyoshi}, {Maihara},
  {Silverman}, {Dalton}, {Lewis}, {Bonfield}, {Lee}, {Lake}, {Macaulay}, \&
  {Clarke}}]{Yabe:11a}
{Yabe}, K., {et~al.} 2011, arXiv:1112.3704

\bibitem[{{Yates} {et~al.}(2012){Yates}, {Kauffmann}, \& {Guo}}]{Yates:12a}
{Yates}, R.~M., {Kauffmann}, G., \& {Guo}, Q. 2012, \mnras, 422, 215

\bibitem[{{Zaritsky} {et~al.}(1994){Zaritsky}, {Kennicutt}, \&
  {Huchra}}]{Zaritsky:94a}
{Zaritsky}, D., {Kennicutt}, Jr., R.~C., \& {Huchra}, J.~P. 1994, \apj, 420, 87

\end{thebibliography}

\end{document}